\documentclass[%
 reprint,
amsmath,
amssymb,
 aps,
 prb,
 floatfix,
]{revtex4-1}

\usepackage[dvips]{graphicx}
\usepackage{dcolumn}
\usepackage{bm}

\usepackage{axodraw4j}
\newcommand{\qel}{Q_e}
\newcommand{\cpl}{c_+}

\newcommand{\bq}{\begin{equation}}
\newcommand{\eq}{\end{equation}}
\newcommand{\bqa}{\begin{eqnarray}}
\newcommand{\eqa}{\end{eqnarray}}
\newcommand{\baa}[1]{\begin{array}{#1}}
\newcommand{\eaa}{\end{array}}
\newcommand{\nll}{\nonumber\\}

\newcommand{\Litwo}{\mbox{${\rm{Li}}_{2}$}}

\newcommand{\ds }{\displaystyle}
\newcommand{\sss}[1]{\scriptscriptstyle{#1}}

\def\mz {M_{\sss{Z}}}

\def\mel{m_e}

\newcommand{\vma}[2]{\delta_{#1}^{#2}}

\newcommand{\stw}{s_{\sss{W}}  }
\newcommand{\ctw}{c_{\sss{W}}  }

\begin{document}

\preprint{APS/}

\title{One-loop Electroweak Radiative Corrections to Polarized Bhabha Scattering}

\author{D. Bardin}\thanks{deceased}
\author{Ya. Dydyshka}%
\author{L. Kalinovskaya}%
\author{L. Rumyantsev}%
\altaffiliation[Also at ]{Institute of Physics, Southern Federal University, Rostov-on-Don,
  344090 Russia}
\author{R. Sadykov}%
\affiliation{%
 Dzhelepov Laboratory of Nuclear Problems, JINR, Dubna, 141980 Russia
}%

\author{A. Arbuzov}
\altaffiliation[Also at ]{Dubna University, Dubna, 141980, Russia}
\author{S. Bondarenko}
 \email{bondarenko@jinr.ru}
\affiliation{
 Bogoliubov Laboratory of Theoretical Physics, JINR, Dubna, 
  141980 Russia
}%

\date{\today}

\begin{abstract}
Theoretical predictions for Bhabha scattering observables are presented 
including complete one-loop electroweak radiative corrections. Longitudinal 
polarization of the initial beams is taken into account.
Numerical results for the asymmetry $A_{LR}$ and the relative correction $\delta$ 
are given for the set of future $e^+e^-$ collider energies $E_{cm}=250, 500, 1000$~GeV 
with various polarization degrees.
\end{abstract}

\maketitle

%
\section{Introduction}
\label{intro}
The complete one-loop electroweak (EW) corrections to unpolarized 
Bhabha scattering~\cite{Bhabha:1936zz} have been thoroughly 
studied for many years in~\cite{Consoli:1979xw} and later in
\cite{Bohm:1984yt,Tobimatsu:1985pp,Bohm:1986fg,Berends:1987jm,Kuroda:1987yi,Bardin:1990xe,Beenakker:1990mb,Beenakker:1990es,Montagna:1993py,Fleischer:2006ht}.
The Bhabha cross section with the  one-loop QED contribution including transverse
and longitudinal polarizations of the incoming beams is presented
in \cite{Hollik:1981bu} and \cite{Hollik:1982wr}. 
Many Monte Carlo event generators for Bhabha scattering were created, see 
e.g.~\cite{Jadach:1996gu} and references therein.
In our review~\cite{Andonov:2004hi} we have presented the {\tt SANC} modules 
for the one-loop electroweak radiative corrections (RC) for Bhabha scattering: 
the Helicity Amplitudes (HA) and Form Factors (FF). 

As compared with hadronic collisions being studied at the LHC, 
$e^{+}e^{-}$ interaction processes have a clean initial state, 
much lower multiplicity,and therefore provide a higher measurement precision
in most cases.
The substantially higher energy range of the future colliders also demands
re-estimation of various effects from both experimental and theoretical sides.  
Precise measurements with polarized beams at the future
$e^ +e^-$ colliders ILC~\cite{homepagesILC} 
and CLIC~\cite{homepagesCLIC} definitely require a modern advanced theoretical 
support~\cite{Khiem:2014cka,Ohl:2006ae,CarloniCalame:2015zev,Riemann:2015zbi}.
In particular, physical programs of the future $e^{+} e^{-}$ linear 
colliders~\cite{Accomando:1997wt,Battaglia:2004mw,Moortgat-Picka:2015yla} 
always demonstrated a great interest to the effects related
to the beam polarization.
 
In the article we present the complete one-loop calculation of 
the EW radiative corrections to Bhabha scattering with polarized beams.
Numerical estimates are shown for the correction 
to the differential distribution in the cosine of the electron scattering 
angle. The relevant contributions to the cross section are calculated 
analytically and then evaluated numerically.

In order to verify our results, we performed several tuned comparisons  
with the results of the alternative systems where available.
The sum of virtual and soft photon Bremsstrahlung contributions 
in the unpolarized case are compared with the {\tt AItalc-1.4} code~\cite{Fleischer:2006ht}.
The polarized Born and hard photon Bremsstrahlung contributions are compared
with the corresponding values obtained with the help of the {\tt WHIZARD} 
code~\cite{Ohl:2006ae,Kilian:2007gr,Kilian:2014nya}.
The unpolarized hard photon contribution is compared analytically
with the result of the {\tt CalcHEP} code~\cite{Belyaev:2012qa}.

This paper is organized as follows.
Sect.~2, the main section of this paper, is fully devoted to 	
the cross section calculation technique at the one-loop level.
Expressions for covariant and helicity amplitudes are presented.
The approach for treatment of polarization effects is discussed.
Sect.~3 contains numerical results for the asymmetry $A_{LR}$ 
and for the relative correction $\delta$ to the differential cross section.
Comparisons with the results of other codes are also presented.
Finally, in Sect.~4 we conclude with a discussion of 
the obtained results.

\section{Differential Cross Section}

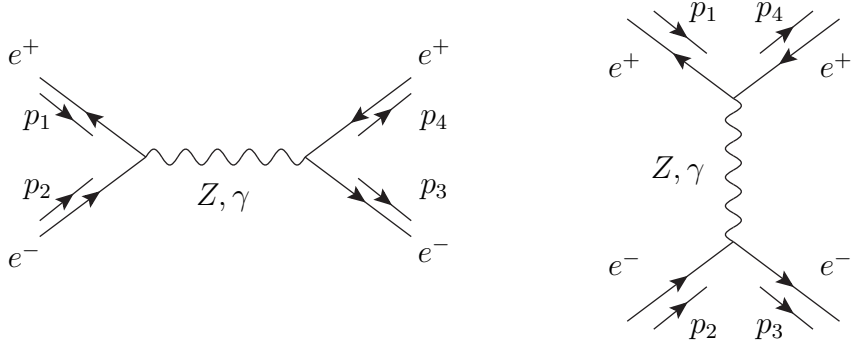
\begin{figure*}
\begin{center}
\begin{tabular}{cc}
   \begin{picture}(125,80)(210,15)
     \Photon(240,40)(300,40){3}{5}
  \Text(270,25)[]{\large $Z,\gamma$}
     \ArrowLine(240,40)(200,70)
     \ArrowLine(200,10)(240,40)
     \ArrowLine(340,70)(300,40)
     \ArrowLine(300,40)(340,10)

     \ArrowLine(200,16)(220,32)
     \ArrowLine(200,64)(220,48)
     \ArrowLine(320,32)(340,16)
     \ArrowLine(320,48)(340,64)

     \Text(200,54)[]{\large $p_1$}
     \Text(200,28)[]{\large $p_2$}
     \Text(350,28)[]{\large $p_3$}
     \Text(350,54)[]{\large $p_4$}

     \Text(195,80)[]{\large$e^+$}
     \Text(195,3)[]{\large $e^-$}
     \Text(350,5)[]{\large $e^-$}
     \Text(350,80)[]{\large$e^+$}
   \end{picture}
\hspace*{10mm}
&
\hspace*{10mm}
  \begin{picture}(125,80)(210,20)
     \Photon(270,67)(270,13){3}{5}
  \Text(250,40)[]{\large $Z,\gamma$}
    \ArrowLine(270,67)(230,97)
    \ArrowLine(310,97)(270,67)
    \ArrowLine(230,-17)(270,13)
    \ArrowLine(270,13)(310,-17)
    \ArrowLine(240,-20)(260,-4)
    \ArrowLine(240,100)(260,84)
    \ArrowLine(280,84)(300,100)
    \ArrowLine(280,-4)(300,-20)
    \Text(230,80)[]{\large $e^+$}
    \Text(230,5)[]{\large $e^-$}
    \Text(310,5)[]{\large $e^-$}
    \Text(310,80)[]{\large $e^+$}
    \Text(260,100)[]{\large $p_1$}
    \Text(285,100)[]{\large $p_4$}
    \Text(285,-20)[]{\large $p_3$}
    \Text(260,-20)[]{\large $p_2$}
  \end{picture}
\end{tabular}
\end{center}
\vspace*{15mm}
\caption[The $s$ (left) and $t$ (right) channels of the Bhabha process at the lowest order.]
        {The $s$ (left) and $t$ (right) channels of the Bhabha process at the lowest order.}
\label{Diagram_schannel}
\end{figure*}

Let us consider scattering of longitudinally polarized $e^+$ and $e^-$ beams 
with the four momenta $p_1$ and $p_2$ for the incoming particles
and $p_3$ and $p_4$ for the outgoing ones: 
\bqa
e^+(p_1)+e^-(p_2) \longrightarrow e^-(p_3)+e^+(p_4).
\label{bornR}
\eqa
Where it is possible, we work in the massless limit and neglect
the effects suppressed by the ratio of the electron mass to the
beam energy. The Feynman diagrams for the Born-level
are shown in Fig.~\ref{Diagram_schannel}.

For the differential cross section $ e^+ e^-\to e^- e^+$ one gets
\begin{equation}
\label{OLsigma}
d\sigma = \frac{1}{32\pi s}\overline{\left|{\cal{A}}\right|^2} d \cos\vartheta\,, 
\end{equation}
where
${\cal{A}}$ is the covariant amplitude (CA) of the process, 
$\sqrt{s}/2$ is the electron energy, and $\vartheta$ is the scattering 
angle in the center-of-mass system (CMS).

\subsection{Covariant one-loop amplitude}

The one-loop covariant amplitude comes out from
the straightforward standard calculation
by means of {\tt SANC} programs 
and procedures of {\em all} diagrams contributing 
to a given process at the tree (Born) and one-loop levels.
The amplitude contains kinematic factors and coupling constants. 
It is parametrized by a certain number 
of form factors (FFs) which are denoted by ${\cal F}$
in general with an index labeling the corresponding Lorentz tensor structure.
The number of FFs is equal to the number of the relevant structures. 
For the processes with non zero tree-level amplitudes the FFs have the form
\bqa
{\cal F} = 1 + k {\tilde{\cal F}}\,,
\eqa 
where ``1'' stands for the Born level and the term ${\tilde{\cal F}}$ with the factor
\bqa
k=\frac{g^2}{16\pi^2}\,,
\label{kaen}
\eqa 
is due to one-loop corrections.
After squaring the amplitude we neglect terms proportional to $k^2$ in order
to get the pure one-loop approximation without any admixture of higher-order
terms which can be added later.

The CA for Bhabha scattering can be written 
(if the electron mass is neglected and because of the symmetry of the process:
${\cal F}_{\sss{LQ}}={\cal F}_{\sss{QL}}$) by the electromagnetic
running coupling constant and four FFs 
with permuted arguments $s$ and $t$ as:
\bqa
{\cal A} &=&{\cal A}_{\gamma}(s)+{\cal A}_{\sss{Z}}(s)
                -\left[{\cal A}_{\gamma}(t)+{\cal A}_{\sss{Z}}(t)\right]
\\
&=& i\,e^2\Biggl\{ \left[ 
 \gamma_\mu \otimes \gamma_\mu \frac{{\cal F}_{\gamma}(s)}{s} 
 -\gamma_\mu \otimes \gamma_\mu \frac{{\cal F}_{\gamma}(t)}{t} \right]
\nll
&+& \frac{\chi_{\sss{Z}}(s)}{s}
 \biggl\{\hspace*{-.3mm} \left(I^{(3)}_{e}\right)^2\hspace*{-1mm}
 \gamma_\mu \gamma_6 \otimes\gamma_\mu \gamma_6 {\cal F}_{\sss{LL}}(s,t,u)
\nll 
&+& 2\delta_{e} I^{(3)}_{e} \gamma_\mu \otimes \gamma_\mu \gamma_6 {\cal F}_{\sss{QL}}(s,t,u)
+\delta^2_{e}\gamma_\mu\otimes\gamma_\mu \,
                   {\cal F}_{\sss{QQ}}(s,t,u)
                   \biggr\}
\nll                   
&-& \frac{\chi_{\sss{Z}}(t)}{t}\biggl\{ s\leftrightarrow t \biggr\}.
\nonumber
\eqa
where $\gamma_6=\left( 1 + \gamma_5 \right)$,
the electron charge $Q_e$, and couplings $~I^{(3)}_e$, ~$ \delta_e = v_e - a_e$.
The symbol $ \otimes $ is used in the following short-hand notation:
\bqa
J^i_\mu \otimes J^j_\mu =
 \bar v(i,p_1) J^i_\mu u(i,p_2)
 \bar u(j,p_3) J^j_\mu v(j,p_4)
\eqa
for $s$ channel and
\bqa
J^i_\mu \otimes J^j_\mu =
 \bar u(i,p_3) J^i_\mu u(i,p_2)
 \bar v(j,p_1) J^j_\mu v(j,p_4)
\eqa
for $t$ channel.
The function $\chi_{\sss{Z}}$ is 
\bqa
\chi_{\sss{Z}}(I)=\frac{1}{4\stw^2\ctw^2}\frac{I}{I-\mz^2+i \mz \Gamma_Z},
\eqa
with $I=s$ or $t$. In the $t$ channel $\Gamma_Z$ is equal to zero. 

\subsection{Virtual, soft, and hard contributions}

The complete result for ${\cal O}(\alpha)$ corrections can be separated into 
the virtual (loop) contribution, the part due to the soft photon emission,
and the last one due to the real hard photon Bremsstrahlung.

$\bullet$ {\bf Born and virtual modules}

Our main approach is to calculate a cross section by squaring non-interfering 
helicity amplitudes (HA). In the CA approach we derive tensor structures and
FFs. The next step is the projection to the helicity basis. As the result one
gets a set of non-interfering amplitudes, since all of them are characterized 
by different sets of helicity quantum numbers.

In this subsection we collect the analytic expressions for the HAs.
There are six non-zero HAs, however, since for Bhabha scattering 
${\cal F}^{\sss Z}_{\sss LQ}={\cal F}^{\sss Z}_{\sss QL}$, the number of independent HAs
is actually reduced to four as expected.

We obtain the compact expression for the Born
(${\cal F}_{{\sss QL},{\sss LL},{\sss QQ}} =1$) 
and the virtual part by HA approach (\ref{VIRT_HA}):
\bqa
\label{VIRT_HA}
{\cal H}_{++++} &=& {\cal H}_{----} =
\\      
&-&2 e^2\,\frac{s}{t} \Bigl[ {\cal F}^{(\gamma,Z)}_{\sss QQ}(t,s,u) 
              - {\chi_{\sss Z}}(t)\vma{e}{} {\cal F}^{\sss Z}_{\sss QL}(t,s,u)\Bigr],
\nll
    {\cal H}_{+-+-} &=& {\cal H}_{-+-+}=
\nll    
&&e^2\,c_{-}
\Bigl[ {\cal F}^{(\gamma,Z)}_{\sss QQ}(s,t,u)-{\chi_{\sss Z}}(s)\vma{e}{}{\cal F}^{\sss Z}_{\sss LQ}(s,t,u)\Bigr],
\nll
    {\cal H}_{+--+} &=&
\nll    
&-&e^2\,\cpl
\Bigl( \Bigl[ {\cal F}^{(\gamma,Z)}_{\sss QQ}(s,t,u)
+{\chi_{\sss Z}}(s)\left( {\cal F}^{\sss Z}_{\sss LL}(s,t,u) \right.
\nll
&-&2\vma{e}{} {\cal F}^{\sss Z}_{\sss LQ}(s,t,u)\left. \right) \Bigr]+\frac{\ds s}{\ds t} \Bigl[ s\leftrightarrow t  \Bigr]
\Bigr),
\nll
{\cal H}_{-++-} &=& - e^2\,\cpl
\Bigl(\left[{\cal F}^{(\gamma,Z)}_{\sss QQ}(s,t,u) \right]
 +\frac{\ds s}{\ds t} \left[  s\leftrightarrow t  \right] \Bigr),
\nonumber
\eqa
where
$c_{\pm}= \cos\vartheta \pm 1$ and
$ {\cal F}^{(\gamma,Z)}_{\sss QQ}(a,b,c)={\cal F}^{(\gamma)}_{\sss QQ}(a,b,c)$
       $+{\chi_{\sss Z}}(a)\vma{e}{2}{\cal F}^{(\sss Z)}_{\sss QQ}(a,b,c)$.


$\bullet$ {\bf Bremsstrahlung module}


The Bremsstrahlung module of the SANC system computes the contributions
due to the soft and inclusive hard real photon emission. 
The soft photon contribution contains infrared divergences and 
has to compensate the corresponding divergences of the one-loop 
virtual QED corrections.

The soft photon Bremsstrahlung correction permits to be calculated analytically. 
It is factorized in front of the Born cross section. 
It depends on the auxiliary parameter which separates kinematical domains
of the soft and hard photon emission in a given reference frame.
The polarization dependence is contained in $\sigma^{\rm Born}$.
The explicit form for the soft photon contribution is
\bqa
 \sigma^{soft} &=& - \sigma^{\rm Born}\frac{\alpha}{\pi}\qel^2  
\Biggl\{
    \left(1+\ln\left(\frac{\mel^2}{s}\right) \right)^2
\\
&&
+  \ln\left(-\frac{u}{s}\right)^2
-\ln\left(-\frac{t}{s}\right)^2 
- 2 \Litwo\left(-\frac{u}{s}\right)
\nll
&&
+ 2 \Litwo\left(-\frac{t}{s}\right)
+ 4 \Litwo\left( 1 \right)-1
\nll
&& + 2\ln\left(\frac{4\omega^2}{\lambda}\right)
    \left[1 + \ln\left(\frac{\mel^2}{s}\right) - \ln\left(\frac{t}{u}\right) \right]
    \Biggr\}.
\nonumber    
\eqa
Here ${\omega}$ is the soft-hard separator and $\lambda$ is an auxiliary infinitesimal 
photon mass and the Spence function $\Litwo(z) = -\int_0^z \dfrac{\ln{(1-x)}}{x}\,dx$. 

The contribution of the hard real photon emission is obtained by direct squaring 
of the matrix element. 
Explicit formulae for the differential distribution of Bhabha process with
hard photon emission are too long to be listed here.

\subsection{Longitudinal polarization}

To study the case of the longitudinal polarization, we produce the helicity amplitudes
and make a formal application of Eq.~(1.15) from~\cite{MoortgatPick:2005cw}.
In our notation the corresponding cross section of Bhabha scattering with
the longitudinal polarization can be expressed as
\bqa
\frac{d\sigma(P_{e^-},P_{e^+})}{d\cos\vartheta}
&=&\frac{1}{128\pi s}  \Bigl[
    (1+P_{e^-})(1+P_{e^+}) \sum_{ij}\lvert{\cal H}_{++ij}\rvert^2
\nll
&&  
  +(1-P_{e^-})(1+P_{e^+}) \sum_{ij}\lvert{\cal H}_{+-ij}\rvert^2
\nll
&&
+(1+P_{e^-})(1-P_{e^+}) \sum_{ij}\lvert{\cal H}_{-+ij}\rvert^2
\nll
&&   +(1-P_{e^-})(1-P_{e^+}) \sum_{ij}\lvert{\cal H}_{--ij}\rvert^2\Bigr].
\label{PolXSec}
\eqa
For the cross check we got analytical zero for the difference 
between the square of the covariant amplitude (we introduced  
the spin density matrix into our procedures) and Eq.~(\ref{PolXSec}).

The asymmetry $A_{LR}$ and the relative correction $\delta$ 
are defined as
\bqa
&&
A_{LR} =
      \frac{d\sigma(-1,1)-d\sigma(1,-1)}{d\sigma(-1,1)+d\sigma(1,-1)},\qquad
\\ \nonumber
&&
\delta =
       \frac{d\sigma^{\text{1-loop}}(P_{e^-},P_{e^+})}{d\sigma^{\text{Born}}(P_{e^-},P_{e^+})}-1,
\eqa
where we omitted $d\cos\vartheta$ for shortness.

\section{Numerical Results and Comparisons \label{NumResultsComp}}

In this section, we present numerical results for EW RC to
Bhabha scattering obtained by means of the {\tt SANC} Monte Carlo 
event generator. Comparisons of our results for specific contributions 
with the ones existing in the literature are also given.

There are many studies devoted to the Bhabha process, 
see e.g.~\cite{Fleischer:2006ht} and references therein. 
It is highly non-trivial to realize a tuned comparison of the numerical results,
since authors usually do not present the complete list of the input parameters
and do not specify the calculation scheme.
Eventually, we compare with the modern packages
{\tt AItalk} and {\tt WHIZARD} which allow to choose 
the following common set of the input parameters:
\begin{eqnarray}
&&\alpha^{-1}(0) = 137.03599976,
\\
&&M_W = 80.4514958 \; \mathrm{GeV}, \quad M_Z = 91.1876 \; \mathrm{GeV},
\nonumber\\
&&\Gamma_Z = 2.49977 \; \mathrm{GeV}, \quad m_e = 0.51099907 \; \mathrm{MeV},
\nonumber\\
&&m_\mu = 0.105658389 \; \mathrm{GeV}, \quad m_\tau = 1.77705 \; \mathrm{GeV},
\nonumber\\
&&m_d = 0.083 \; \mathrm{GeV}, \quad m_s = 0.215 \; \mathrm{GeV},
\nonumber\\
&&m_b = 4.7 \; \mathrm{GeV}, \quad m_u = 0.062 \; \mathrm{GeV},
\nonumber\\
&&m_c = 1.5 \; \mathrm{GeV}, \quad m_t = 173.8 \; \mathrm{GeV}.
\nonumber
\end{eqnarray}
The $\alpha(0)$ EW scheme is used in all calculations.

All results are obtained for the set of the energy $E_{cm}=250$, $500$, and $1000$~GeV
for the following magnitudes of the electron $(P_{e^-})$ and the positron $(P_{e^+})$
beam polarizations: $(0,0)$, $(-0.8,0)$, $(-0.8,-0.6)$, $(-0.8,0.6)$.

$\bullet$ {\bf Comparison for Born and hard photon contributions}

First of all we verified an agreement between our analytic result for 
the unpolarized hard photon Bremsstrahlung 
process cross section with the one obtained with the help 
of the {\tt CalcHEP} system~\cite{Belyaev:2012qa}.

The numerical comparison for the hard photon Brem\-sstrahlung cross section 
with polarized initial particles is performed using the {\tt WHIZARD} system.
Table~\ref{sanc_vs_whiz_hard} shows the good agreement between 
the {\tt SANC} results (the second row) for the Born and hard photon Bremsstrahlung 
contributions with the ones obtained with the help of the {\tt WHIZARD} (the first row)
program~\cite{Ohl:2000pr}. 
The range of scattering angles for the final electrons and positrons in this comparison
was limited by the condition $|\cos\theta|<0.9$ with 
the additional condition $E_{\gamma} \geq 1$ GeV for the emitted photon energy. 

\begin{table}[b]
\caption{Tuned comparison of the {\tt SANC} and {\tt WHIZARD} results for the Born and 
hard Bremsstrahlung contributions to polarized Bhabha scattering
for $\sqrt{s} = 250$, $500$, and $1000$~GeV.}
\label{sanc_vs_whiz_hard}
\begin{ruledtabular}
\begin{tabular}{lcccc}
$P_{e^-}$, $P_{e^+}$ & 0, 0 & -0.8, 0 & -0.8, -0.6 & -0.8, 0.6\\
\hline
\multicolumn{5}{c}{$\sqrt{s} = 250$ GeV}\\
\hline
$\sigma^{\text{Born}}$, pb & 56.677(1) & 57.774(1) & 56.272(1) & 59.276(1)\\
$\sigma^{\text{Born}}$, pb & 56.677(1) & 57.775(1) & 56.272(1) & 59.275(1)\\
$\sigma^{\text{hard}}$, pb  & 48.62(1) & 49.58(1) & 48.74(1) & 50.40(1)\\
$\sigma^{\text{hard}}$, pb  & 48.65(1) & 49.56(1) & 48.78(1) & 50.44(1)\\
\hline
\multicolumn{5}{c}{$\sqrt{s} = 500$ GeV}\\
\hline
$\sigma^{\text{Born}}$, pb  & 14.379(1) & 15.030(1) & 12.706(1) & 17.355(1)\\
$\sigma^{\text{Born}}$, pb  & 14.379(1) & 15.030(1) & 12.706(1) & 17.354(1)\\
$\sigma^{\text{hard}}$, pb  & 15.14(1)  & 15.81(1) & 13.54(1) & 18.07(1)\\
$\sigma^{\text{hard}}$, pb  & 15.12(1)  & 15.79(1) & 13.55(1) & 18.11(2)\\
\hline
\multicolumn{5}{c}{$\sqrt{s} = 1000$ GeV}\\
\hline
$\sigma^{\text{Born}}$, pb  & 3.6792(1) & 3.9057(1) & 3.0358(1) & 4.7756(1)\\
$\sigma^{\text{Born}}$, pb  & 3.6792(1) & 3.9057(1) & 3.0358(1) & 4.7755(1)\\
$\sigma^{\text{hard}}$, pb  & 4.693(1)  & 4.976(1)  & 3.912(1)  & 6.041(1)\\
$\sigma^{\text{hard}}$, pb  & 4.694(1)  & 4.975(1)  & 3.913(1)  & 6.043(2)\\
\end{tabular}
\end{ruledtabular}
\end{table}

$\bullet$ {\bf  Comparison of virtual and soft photon\\ Bremsstrahlung contributions}

We obtained a very good agreement (six significant digits) in the comparison 
of the {\tt SANC} and {\tt AItalc-1.4}~\cite{Fleischer:2006ht} results for 
the unpolarized differential Born cross section and for the sum of 
the virtual and the soft photon Bremsstrahlung contributions. 
The comparison was done for the different values of the
scattering angles $(\cos \vartheta$: from $-0.9$ up to $+0.9999)$.

$\bullet$ {\bf Results for Born and 1-loop cross section}

\begin{table}[!ht]
\caption{Born and 1-loop cross sections of Bhabha scattering and the corresponding 
relative corrections $\delta$ for $\sqrt{s} = 250$, $500$ and $1000$~GeV.}
\label{Table:sanc_delta}
\begin{ruledtabular}
\begin{tabular}{lcccc}
$P_{e^-}$, $P_{e^+}$ & 0, 0 & -0.8, 0 & -0.8, -0.6 & -0.8, 0.6\\
\hline
\multicolumn{5}{c}{$\sqrt{s} = 250$ GeV}\\
\hline
$\sigma_{e^+e^-}^{\text{Born}}$, pb & 56.6763(1) & 57.7738(1) & 56.2725(4) & 59.2753(5)\\
$\sigma_{e^+e^-}^{\text{1-loop}}$, pb & 61.731(6) & 62.587(6) & 61.878(6) & 63.287(7)\\
$\delta$, \% & 8.92(1) & 8.33(1) & 9.96(1) & 6.77(1) \\
\hline
\multicolumn{5}{c}{$\sqrt{s} = 500$ GeV}\\
\hline
$\sigma_{e^+e^-}^{\text{Born}}$, pb & 14.3789(1) & 15.0305(1) & 12.7061(1) & 17.3550(2)\\
$\sigma_{e^+e^-}^{\text{1-loop}}$, pb & 15.465(2) & 15.870(2) & 13.861(1) & 17.884(2)\\
$\delta$, \% & 7.56(1) & 5.59(1) & 9.09(1) & 3.05(1)\\
\hline
\multicolumn{5}{c}{$\sqrt{s} = 1000$ GeV}\\
\hline
$\sigma_{e^+e^-}^{\text{Born}}$, pb & 3.67921(1) & 3.90568(1) & 3.03577(3) & 4.77562(5)\\
$\sigma_{e^+e^-}^{\text{1-loop}}$, pb & 3.8637(4) & 3.9445(4) & 3.2332(3) & 4.6542(7)\\
$\delta$, \% & 5.02(1) & 0.99(1) & 6.50(1) & -2.54(1)\\
\end{tabular}
\end{ruledtabular}
\end{table}

\begin{figure*}
\[
\begin{array}{ccc}
  \includegraphics[width=80mm]{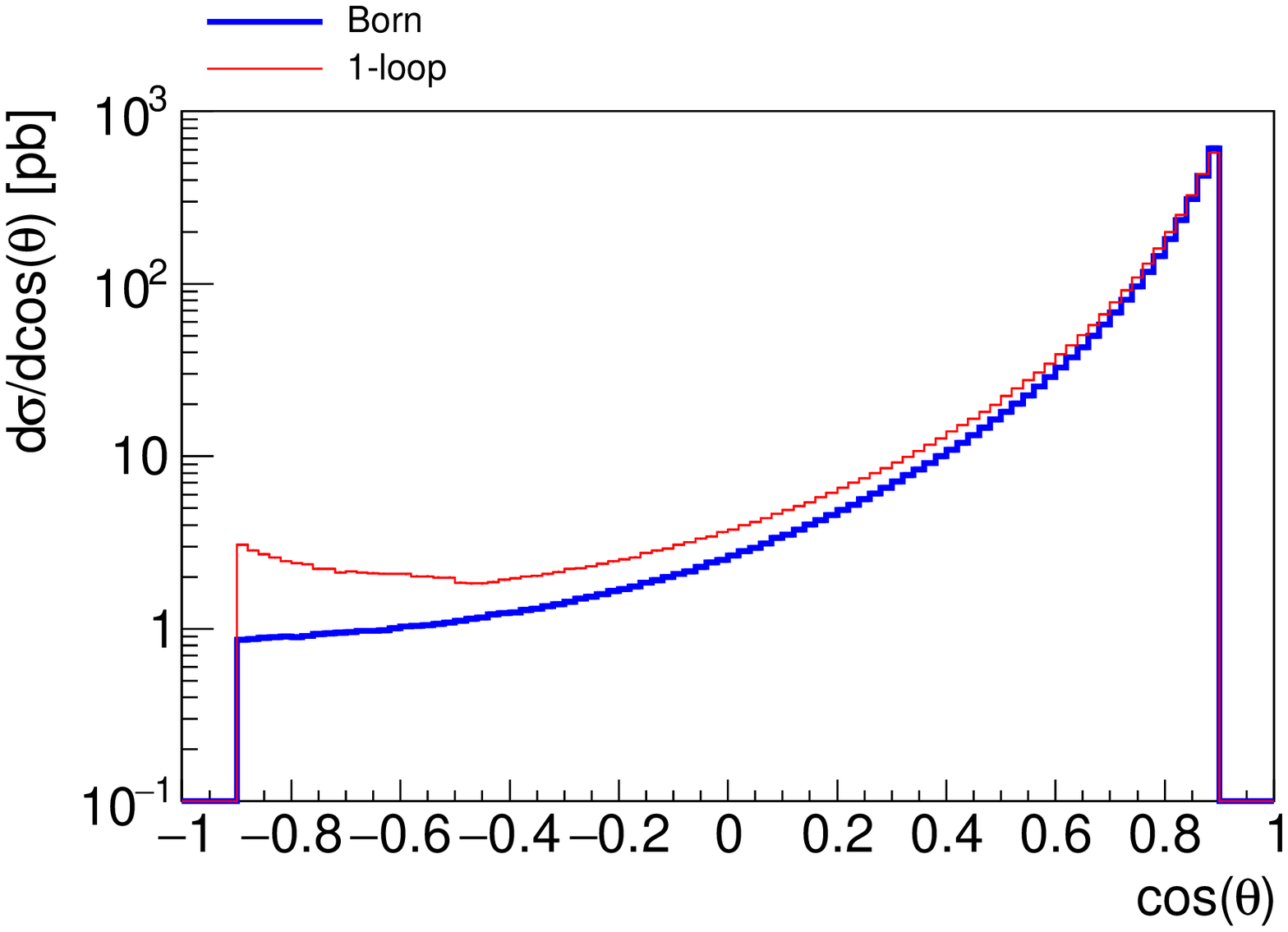}
  &&
  \includegraphics[width=80mm]{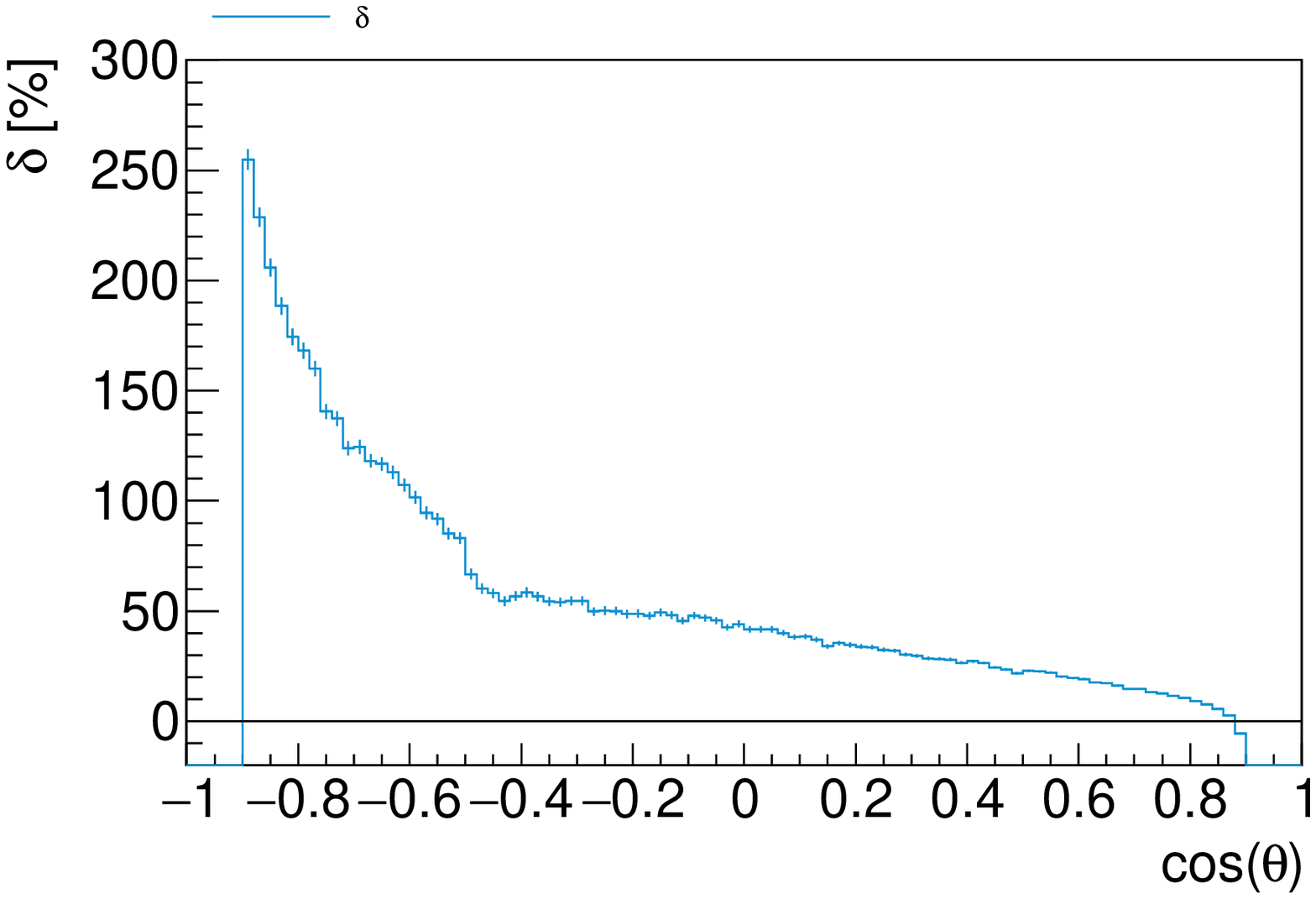}
  \end{array}
\]
\caption{The differential cross section (left) [in pb] and 
the relative correction $\delta$ (right) [in \%] vs. the cosine 
of the electron scattering angle for $\sqrt{s}=250$~GeV.}
\label{el_delta_costh_250}
\end{figure*}

\begin{figure*}
\[
\begin{array}{ccc}
  \includegraphics[width=80mm]{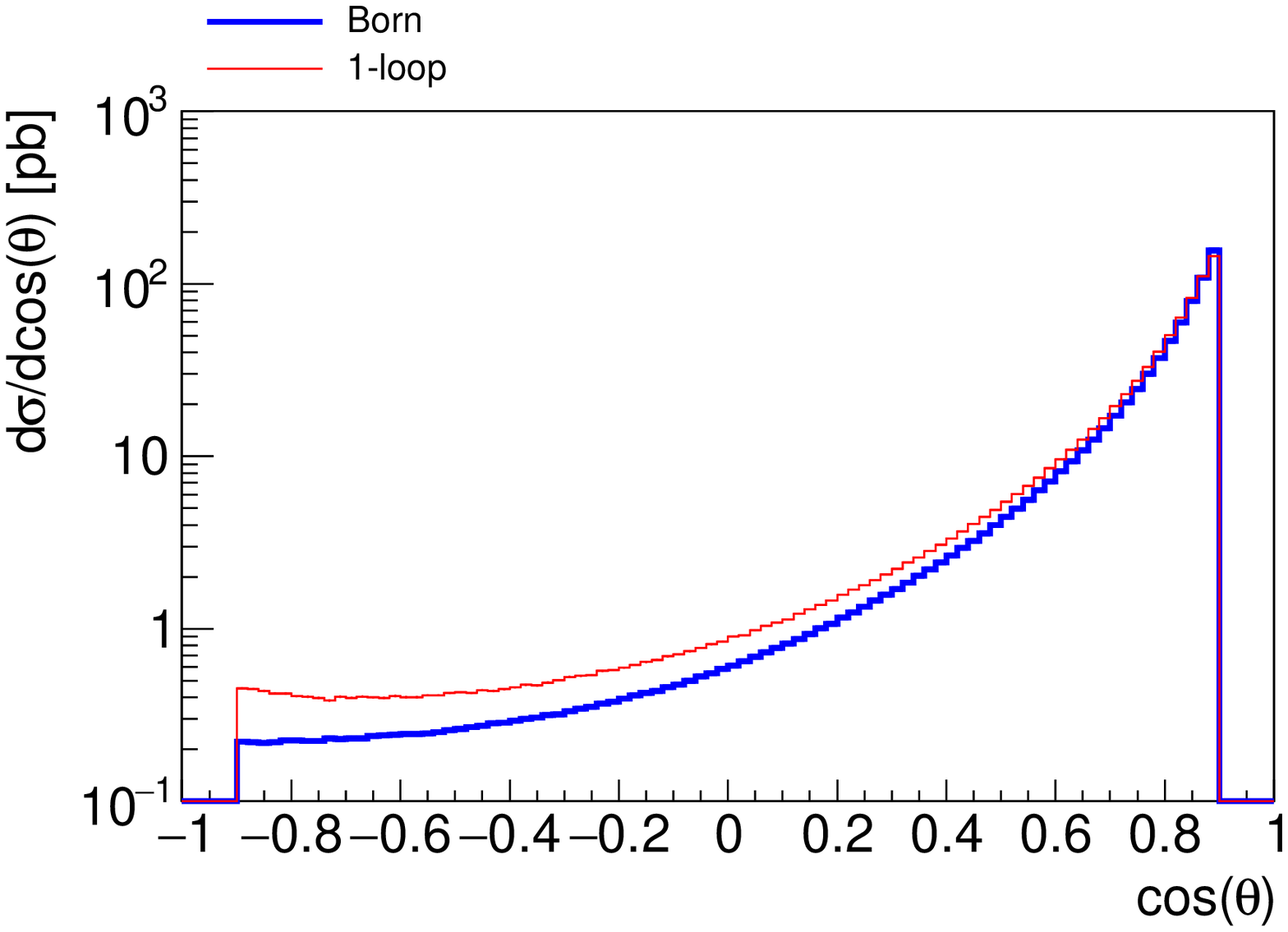}
  &&
  \includegraphics[width=80mm]{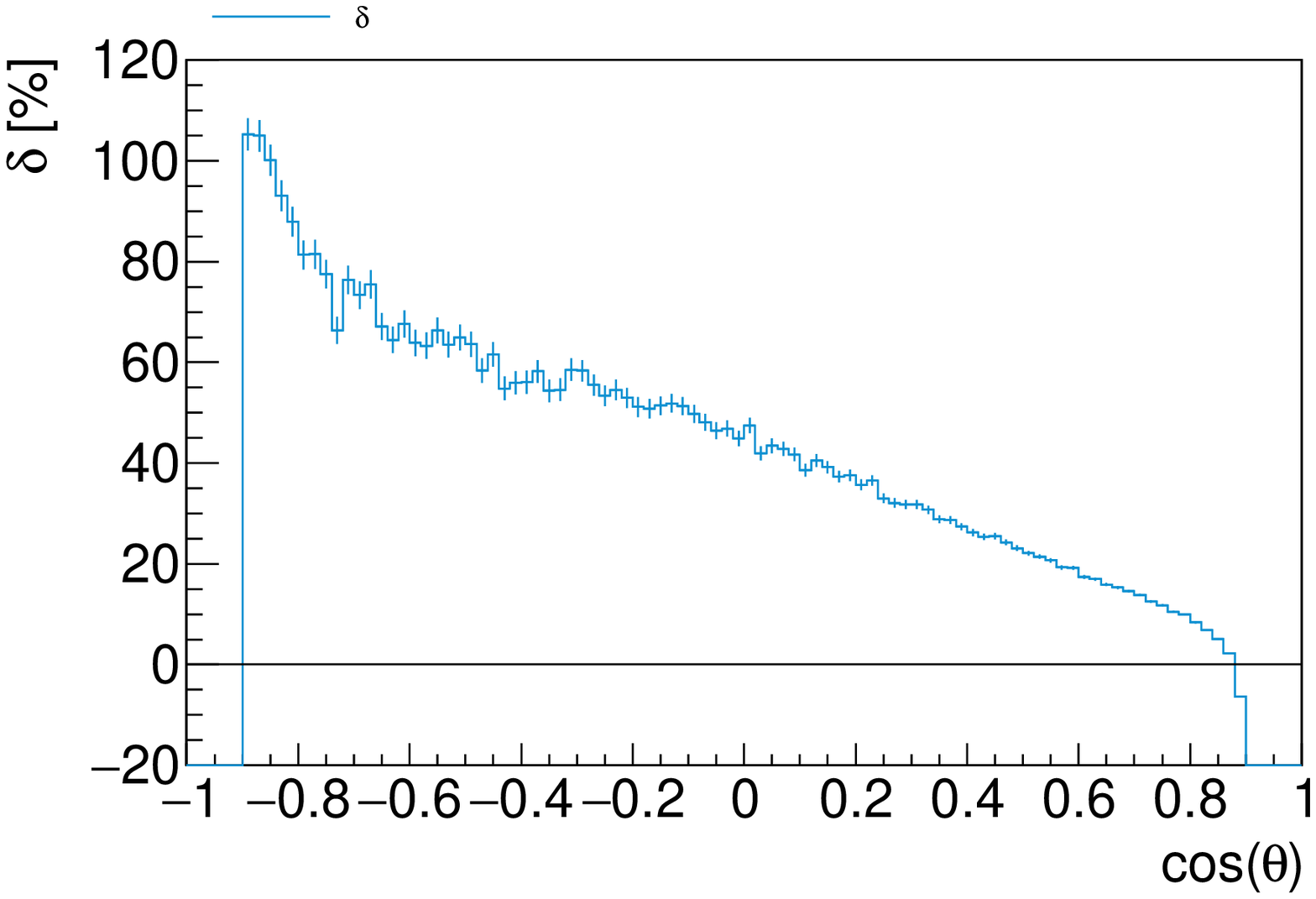}
  \end{array}
\]
\caption{The differential cross section (left) [in pb] and 
the relative correction $\delta$ (right) [in \%] vs. the cosine 
of the electron scattering angle for $\sqrt{s}=500$~GeV.}
\label{el_delta_costh_500}
\end{figure*}

\begin{figure*}
\[
\begin{array}{ccc}
  \includegraphics[width=80mm]{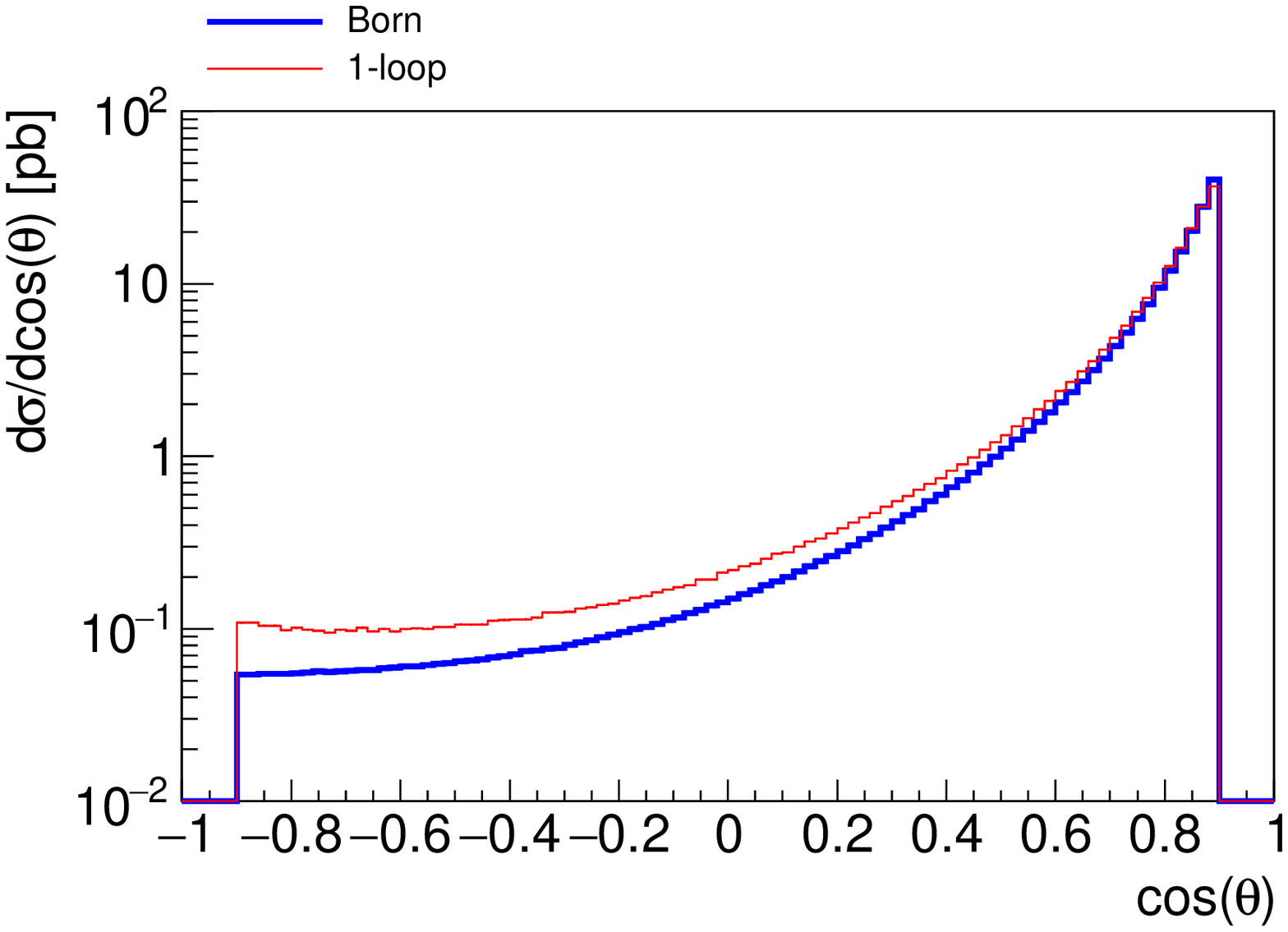}
  &&
  \includegraphics[width=80mm]{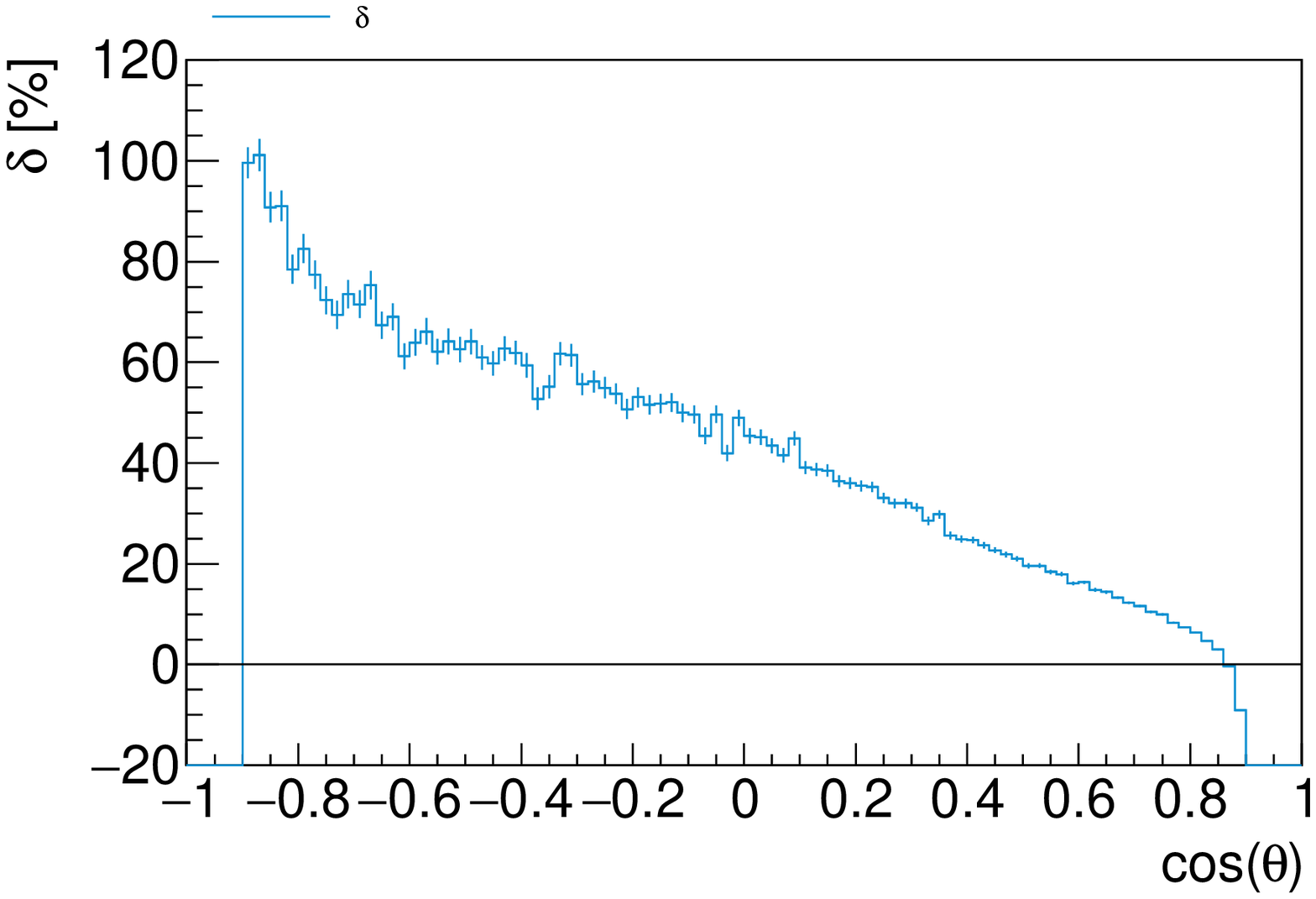}
  \end{array}
\]
\caption{The differential cross section (left) [in pb] and 
the relative correction $\delta$ (right) [in \%] vs. the cosine 
of the electron scattering angle for $\sqrt{s}=1000$~GeV.}
\label{el_delta_costh_1000}
\end{figure*}

\begin{figure}
  \includegraphics[width=80mm]{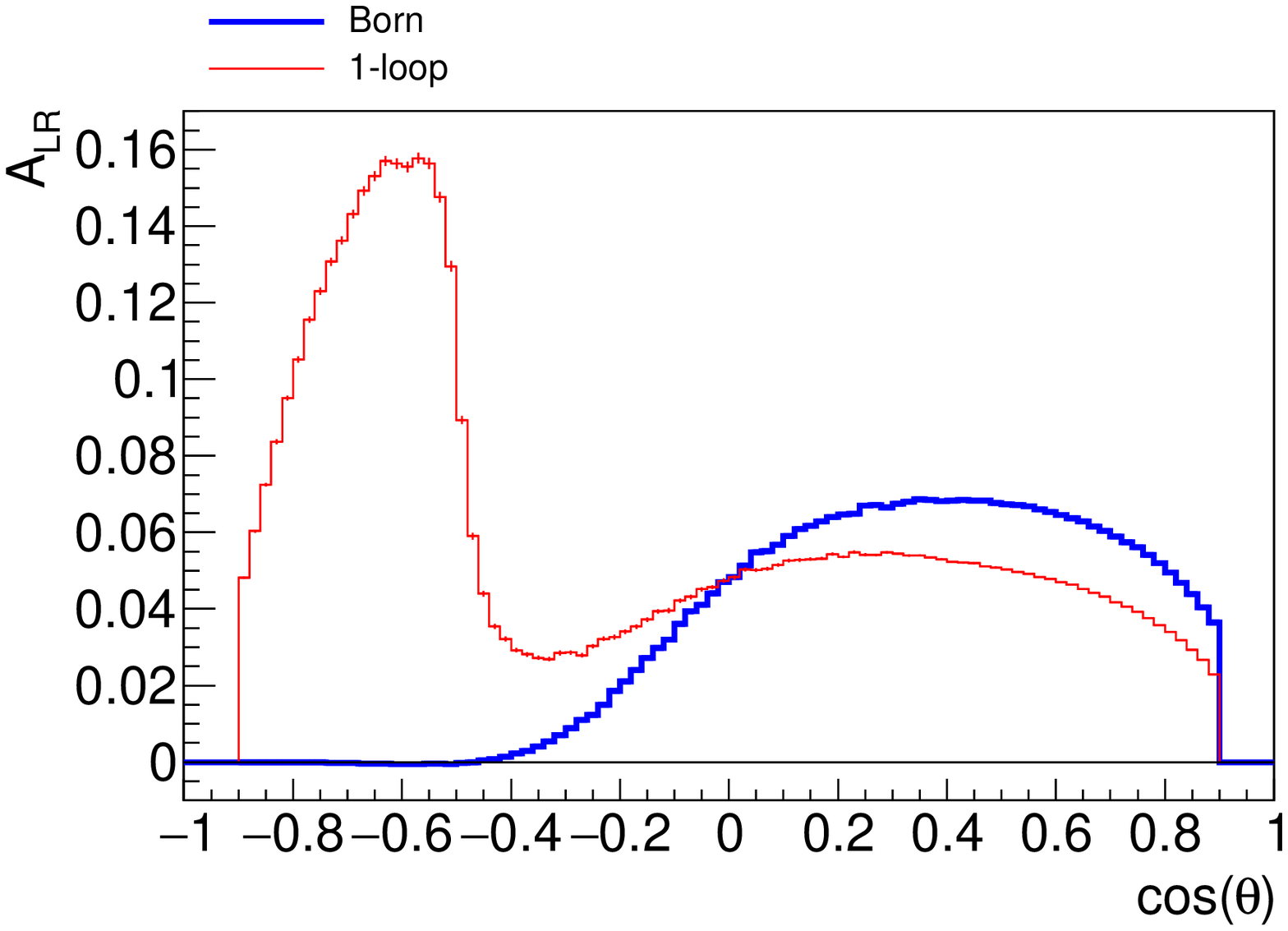}
\caption{The left-right asymmetry $A_{LR}$ as a function of the cosine of the electron 
scattering angle at $\sqrt{s}=250$~GeV.}
\label{a_el_costh-250}
\end{figure}
\begin{figure}
  \includegraphics[width=80mm]{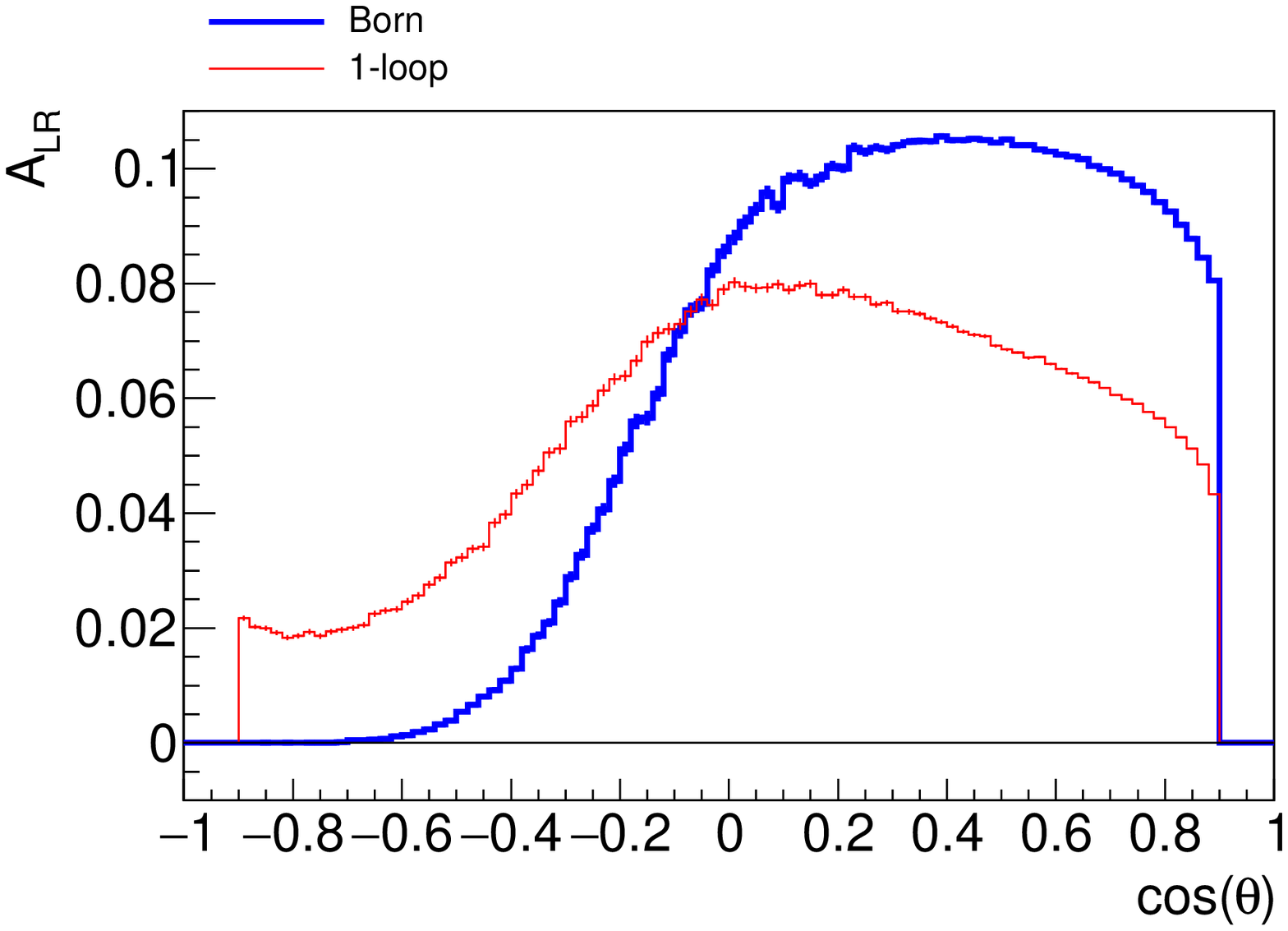}
\caption{The left-right asymmetry $A_{LR}$ as a function of the cosine of the electron 
scattering angle at $\sqrt{s}=500$~GeV.}
\label{a_el_costh-500}
\end{figure}
\begin{figure}
  \includegraphics[width=80mm]{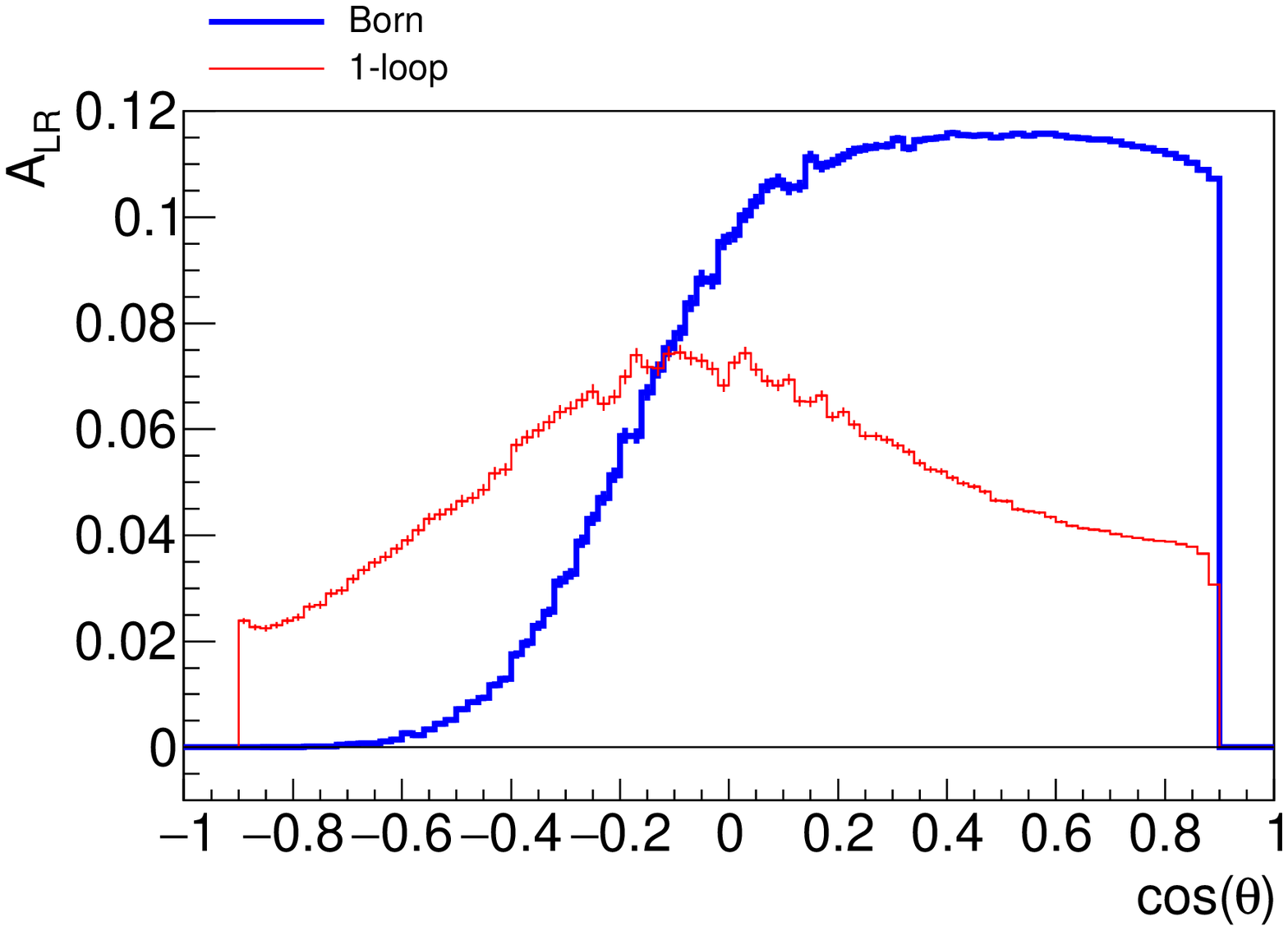}  
\caption{The left-right asymmetry $A_{LR}$ as a function of the cosine of the electron 
scattering angle at $\sqrt{s}=1000$~GeV.}
\label{a_el_costh-1000}
\end{figure}

The unpolarized differential cross section of Bhabha scattering
and the relative ${\mathcal{O}}(\alpha)$ 
correction $\delta$ (in percent) as a function
of the electron scattering angle are shown in Figs.~\ref{el_delta_costh_250},
\ref{el_delta_costh_500}, and \ref{el_delta_costh_1000} for $|\cos\theta|<0.9$
and different CMS energies.

The huge relative radiative corrections for the
backward scattering angles are due to the smallness of the Born cross section 
in this domain, that does not mean any problem with the perturbation theory.

The integrated cross section of the Bhabha scattering and the relative
correction $\delta$ are given in the Table~\ref{Table:sanc_delta} for
various energies and beam polarization degrees.

The $A_{LR}$ asymmetry at $\sqrt{s}=250$, $500$, and $1000$~GeV
is shown in Figs.~\ref{a_el_costh-250}-\ref{a_el_costh-1000}. 
One can see that the EW radiative corrections affect the asymmetry very strongly.

\section{Conclusions}

The theoretical description of Bhabha scattering with taking into account the radiative 
corrections is crucial for the high-precision measurement of this process 
and thus for luminosity monitoring at the modern and future $e^+e^-$ colliders.
Consideration of the beam polarization is a novel requirement
for the theoretical predictions for the $e^+e^-$ collisions at the energies of CLIC and ILC.
Moreover our results can be relevant for the physical program of the 
Super $c-\tau$ factory~\cite{Eidelman:2015wja} planned to be built in Novosibirsk,
where polarization of the electron beam is also foreseen.

We show that the complete ${\mathcal{O}}(\alpha)$ electroweak radiative corrections 
provide a considerable impact on the differential cross section and the
left-right asymmetry. Moreover, the corrections themselves are rather sensitive 
to polarization degrees of the initial beams.

The observed  magnitude of the first order corrections certainly 
provokes the question about the size of the higher-order corrections to this process.
Some of those corrections are known in the literature (but mainly the pure QED ones 
for the unpolarized case). To estimate the theoretical uncertainty we plan to implement
the known higher-order corrections to Bhabha scattering into our Monte Carlo 
event generator.

\begin{acknowledgments}
We express our sincere gratitude to Sabine Riemann~\cite{ReportSR} for fruitful 
stimulating discussions.
We thank Gizo Nanava for collaboration on the {\tt SANC} system.
Results were obtained within the framework of State Program 
No. 3.9696.2017/8.9 from the Ministry of Education and Science of Russia.
Authors are grateful to Dr. A.~Gladyshev for the help in preparation of the manuscript.
\end{acknowledgments}


\end{document}